\documentclass{article}
\usepackage[utf8]{inputenc}
\usepackage[T1]{fontenc}
\usepackage[english]{babel}
\usepackage{amsmath}
\usepackage{amssymb,amsfonts,textcomp}
\usepackage{hyperref}
\hypersetup{ colorlinks=true, linkcolor=blue, citecolor=blue, filecolor=blue, urlcolor=blue, pdftitle=No entailing laws but enablement in the evolution of the biosphere', pdfauthor=, pdfsubject=, pdfkeywords=}
\usepackage{geometry}
\title{No entailing laws, but enablement in the evolution of the biosphere}
\author{Giuseppe Longo \and Maël Montévil \and Stuart Kauffman\footnote{Authors affliations: GL and MM: CNRS, CREA - Polytechnique et CIRPHLESS - Ecole Normale Sup., Paris (Fr.); SK:  Tampere University of Technology (Fi), University of Vermont (USA), Calgary University, Canada; longo@ens.fr, montevil@di.ens.fr, stukauffman@gmail.com.}}

\begin{document}
\maketitle
\begin{abstract}
Biological evolution is a complex blend of ever changing structural stability, variability and emergence of new phenotypes, niches, ecosystems.  We wish to argue that the evolution of life marks the end of a physics world view of law entailed dynamics. Our considerations depend upon discussing the variability of the very ''contexts of life'': the interactions between organisms, biological niches and ecosystems. These are ever changing, intrinsically indeterminate and even unprestatable: we do not know ahead of time the "niches" which constitute the boundary conditions on selection. More generally, by the mathematical unprestatability of the "phase space" (space of possibilities), no laws of motion can be formulated for evolution.  We call this radical emergence, from life to life. The purpose of this paper is the integration of variation and diversity in a sound conceptual frame and situate unpredictability at a novel theoretical level, that of the very phase space.

Our argument will be carried on in close comparisons with physics and the mathematical constructions of phase spaces in that discipline. The role of (theoretical) symmetries as invariant preserving transformations will allow us to understand the nature of physical phase spaces and to stress the differences required for a sound biological theoretizing. In this frame, we discuss the novel notion of ''enablement". Life lives in a web of enablement and radical emergence.  This will restrict causal analyses  to differential cases (a difference that causes a difference). Mutations or other causal differences will allow  us to stress that ''non conservation principles" are at the core of evolution, in contrast to physical dynamics, largely based on conservation principles as symmetries. Critical transitions, the main locus of symmetry changes in physics, will be discussed, and lead to  ''extended criticality" as a conceptual frame for a better understanding of the living state of matter.
\end{abstract}

\section{Overview}
The aim of this article is to demonstrate that the mode of understanding in physics since Newton, namely differential equations, initial and boundary conditions, then integration which constitutes deduction, which in turn constitutes “entailment”, fails fundamentally for the evolution of life. No law in the physical sense, we will argue, entails the evolution of life. If we are correct, this spells the end of “strong reductionism”, the long held belief that a set of laws “down there” entails all that happens in the universe. More, if no law entails the evolution of life, yet the biosphere is the most complex system we know of in the universe, it has managed to come into existence without an entailing law. Then such law is not necessary for extraordinary complexity to arise and thrive. We need new ways to think about how current life organization can have come into being and persists. Those ways include coming to understand how what we will call organisms as “Kantian wholes” co-create their worlds with one 
another.

The heart of our considerations are these: 1) In physics, we can prestate the configuration space or \emph{phase space}, a crucial notion in this paper. Dynamics are geodetics within such prestated phase spaces, which may be very abstract, like in Quantum Mechanics. 2) In biological evolution, the phase space itself changes persistently. More it does so in ways that cannot be prestated. 3) Because we cannot prestate the ever changing phase space of biological evolution, we have no settled relations by which we can write down the “equations of motion” of the ever new biologically “relevant observables and parameters” revealed after the fact by selection acting on Kantian wholes in biological evolution, but that we cannot prestate. More, we cannot prestate the adaptive “niche” as a boundary condition, so could not integrate the equations of motion even were we to have them. 4) If the above is true, no law entails the evolution of the biosphere. 5) If by “cause”, we mean what gives a differential effect \emph{
entailed by law}, then we can assign no cause in the “diachronic” evolution of the biosphere. 6) In place of “cause” in this diachronic evolution, we will find “enablement”, ie making possible -- a key notion in our analysis. 7) Our thesis does not obviate reductive explanations of organisms as synchronic entities, such as an ultimate physical account of the behavior of an existing heart, once evolved\footnote{The intrinsic unpredictability (or non-pre-definability) of the biological phase space was first hinted in \cite{kauffman2000investigations} and, on different grounds in \cite{bailly2011} and in the orginal version, in French, Hermann, 2006; see also \cite{kauffman2008,bailly2008}.}.

Our analysis is centered on cells and organisms as Kantian wholes, where the whole exists for and by means of the parts, and the parts for and by means of the whole. Given a Kantian whole, the “function” of a part in sustaining the whole is definable. Other synchronic causal consequences are irrelevant side effects. An essential feature of our analysis is that at levels of complexity above the atom, for example for molecules, the universe is grossly non-ergodic, that is it does not explore all possible paths or configurations. We will not make all possible proteins length 200 amino acids in 10 to the 39th times the lifetime of the universe, even were all 10 to the 80th particles making such proteins on the Planck time scale. Thus, the existence in the universe of a heart, which Darwin tells us is due to its selective advantage in a sequence of descendant, Kantian whole organisms, is physically important: Most complex things will not ever exist. Thus Darwin's theory is telling us how hearts exist in the 
universe. Kantian wholes, married to self reproduction and Darwinian evolution, are part of the non-ergodic, historical becoming of the universe, and, we claim, beyond entailing law. A deep aspect of the freedom from entailing law in the evolution of organisms is that the possible “uses” of a given part or process of an organism are, both indefinite and unoderable, in our views, thus, a fortiori, no effective procedure, or algorithm can list them.  Thus when selection acts at the level of the whole organism, we cannot have pre-listed the newly relevant functional features of its parts revealed by selection.  It follows that we cannot prestate the ever changing relevant observables and variables revealed by selection, so --- our main theme --- cannot write equations of motion for the evolving biopshere. Nor can we in general prestate the boundary conditions on selection, i.e. the “niche”, so we could not integrate the equations of motion that we do not have anyway. In short, no law entails the evolution of 
the 
biosphere, nor, more specifically, of an organism phylogenetic trajectory. Moreover, we claim, niches and ecosystems ''enable" the formation of life, and causal relations should be seen only in differential effects.

The key argument will be given in reference and contrast to the role of symmetries and conservation laws in physics. We will show that symmetries and the mathematics of invariants and invariant preserving transformations cannot be transferred as such to suitable theoretical frames for biology, in particular to analyses of biological evolution. We will note that the construction of phase spaces for physics has been largely, or even exclusively, based on invariant properties of ''trajectories", ie on symmetries. This fails in biological theoretizing, since phylogenetic trajectories may be view as continual symmetry changes. In our perspective, These continual symmetry changes are correlated to unprestatable changes of the state space itself. Of course, the key point, extensively stressed below, is the proposal of the suitable observables in biology. These are dictated by the chosen theory. For us, this is Darwin's Evolution.

In this theoretical frame, we cannot prestate or list the possible selective biological environments, in view of the provable impossibility to prestate the intended ''coherence structures" and their symmetries. As a consequence, the set of objects or processes able to carry out a given use, or Kantian function, is also indefinite and unorderable, so not enumerable. But such ever novel parts and processes as adaptations and Darwinian preadaptations arise all the time, often by quantum indeterminate, acausal, random mutations, and find a use, and often a novel function as parts in the organism in an unprestatable selective environment, so are grafted into the organism, thereby changing the phase space of evolution, we will argue, in an unprestatable, incompressible, way. Concurrently, we note that we can think of a reproducing cell or organism as achieving a “task closure” in some set of tasks, such as mitosis, the behavior of chemosmotic pumps and so on. But this task closure is achieved via the biotic and 
abiotic environment. But only selection acting at the level of the Kantian wholes, reveals after the fact, the newly relevant features of the organism and the environment that constitute the task closure and the new “niche” of the now surviving organism. In short, the organism and its niche, are co-constituted in a circular way that cannot be prestated. Again, we will argue that no laws of motion, nor boundary condition to integrate such laws were we to have them, can be formulated. No law entails the evolution of the biosphere.

Finally, and stunningly, evolution creates, without selection {\it acting to do so}, new “adjacent possible empty niches” which enable new possible directions of evolution. This is radical emergence from life to life. Further, the evolution of a new organism to live in a new “adjacent possible empty niche”, often arises due to one or a sequence of quantum events, at the molecular level, which are acausal. Thus the niche does not cause, but “enables” the radical emergence. Not only is life caught in a web of causes, it is part of a co-constituting web of enablement and radical emergence.

If correct, reductionism reaches a terminus at the watershed of life. With Heraclitus we say of life: The world bubbles forth.

\section{ Physical phase spaces.}

In order to understand the specific difficulties of biological theoretizing, we will first shortly recall the role of “phase spaces” in physical theories.

In physics, the phase space is given by the pertinent observables and parameters. As we will recall below, these may be, for example, the momentum and position, or energy and time. In Hamiltonian classical mechanics and in Quantum Physics, these observables and variables happen to be “conjugated”, an expression of their tight relation and pertinence. These mathematical spaces are the spaces in which the trajectories are determined: even in Quantum Physics, when taking Hilbert’s spaces as phase spaces for the wave function, Schrödinger’s equation \emph{determines} the dynamics of a probability density and the indeterministic aspect of quantum mechanics appears when quantum measurement projects the state vector.

As a matter of fact, one of the major challenges for a (theoretical) physicist is to invent the right (pertinent) spaces or phase space. Well before the invention of the notion of “phase space”, Newton’s analysis of trajectories was fully embedded in Descartes spaces, a “condition of possibility”, Kant will explain, for physics to be done. By this, Newton unified (he did not reduce) Galileo’s analysis of falling bodies, including apples, to planetary orbits. Then Newton derived Kepler’s ellipsis of a planet around the Sun from his equations; this is the astonishing birth of modern mathematical-physics as capable of predicting exactly a trajectory inside the theory. But, since Poincaré, we know that if the planets around the Sun are two, prediction is impossible due to determinsitic chaos. Their trajectories are fully determined by Newton-Laplace equations; but in this case, though, their non-linearity yields the absence almost everywhere of analytic solutions and forbids predictability, even along well 
determined trajectories at equilibrium, in perfectly pre-defined phase spaces.

Poincaré’s analysis of chaotic dynamics was essentially based on his invention of the so-called Poincaré section (analyze planetary orbits only by their crossing a given plane) and of the use of momentum as a key observable. In his analysis of chaoticity, stable and unstable trajectories in the position-momentum phase space, nearly intersect infinitely often, in “infinitely tight meshes” and are also “folded upon themselves without ever intersecting themselves”, (1892).

As for thermodynamics, Boyle, Carnot and Gay Lussac, decided to focus on pressure, volume and temperature, as the relevant observables: the phase space\footnote{The term phase space is often restricted to a position/momentum space; we use it in this text in the general sense of the suitable/intended space of the mathematical and/or theoretical description of the system. } was chosen in view of its pertinence, totally disregarding the familiar fact that gases are made out of particles. Boltzmann later unified the principles of thermodynamic to a particle’s viewpoint and later to Newtonian trajectories by the ergodic hypothesis: no reduction to Newtonian trajectories, rather a unification at the infinite time limit of the thermodynamic integral, under the novel assumption of “molecular chaos”, an “asymptotic” unification, rather extraneous to the Newton-Laplace theory of movement.

We already mentioned Schrödinger’s invention of a suitable phase space for the state function in Quantum Mechanics: the very abstract Hilbert space of probability densities, very far form ordinary space-time.

As a matter of fact, we claim that these conceptual/mathematical constructions of a pre-given phase space do not apply to biology and we will motivate this by different levels of analysis. First we will discuss naive constructions of a microscopic phase space in biology, then we will show that evolution leads to continual unprestatable changes in phase spaces and finally we will analyse the role of symmetries and conservation properties in physics, which are, instead, continually broken or changed in biological dynamics. This yields the impossibility of proposing a pre-given biologically pertinent phase space, which would accommodate all possible phylogenetic trajectories. Therefore and most critically, we cannot write equations of motion for the evolving biosphere, nor prestate the niche boundary conditions,  which therefore does not allow us to integrate the equations of motion we do not have anyway.

\section{Biology and microphysical descriptions: non-ergodicity and quantum effects.}

There are some immediate difficulties with the “naive” approach to the construction of a phase space in biology. We will discuss the question of ergodicity and the combination of quantum and classical phenomena in evolution.

As mentioned in the overview, an easy combinatorial argument shows that even at the TIME scale of the Universe, all possible macromolecules, or even proteins of length 200 amino acids cannot be explored. So, their “composition” in a new organ or organism (thus, in a phenotype) cannot be the result of the ergodicity of physical dynamics\footnote{Notice here that this argument only states that ergodicity in the molecular phase space does not help to understand the biological dynamics. The argument allows trajectories to be ergodic\emph{ in infinite time}. We can then say that ergodicity is biologically irrelevant and can take this irrelevance as a principle. }. Because of this non-ergodicity, history enters. More, 
most complex things will never exist. Later, in discussing Kantian wholes where the parts exist for and by means of the whole and the whole for and by means of the parts, the physical relevance is that in the non-ergodic universe, hearts and humans, via evolution, do exist in the universe.

We should first compare this situation with the role of ergodicity in statistical mechanics. The basic assumption of (most) statistical mechanics is a symmetry between states with the same energetic level, which allows us to analyze their probabilities (on the relevant time scales). This assumption is grounded on an hypothesis of ergodicity as for the dynamics of the particles: in the limit of infinite time, they “go everywhere” in the intended phase space. In this case, the situation is described on the basis of energetic considerations (energy conservation properties, typically), without having to take into account the Newtonian trajectory or the history of the system.

In biology, the non-ergodicity in the molecular phase space allows us to argue that the dynamic cannot be described without temporal considerations, even when considering only the molecular aspects of biological systems, let alone morphological and other aspects. In other terms, it means that the relevant symmetries also in the tentative phase space for molecules depend on history, in contrast to statistical mechanics.

Note that some cases of non-ergodicity are well studied in physics. Symmetry breaking phase transitions is a simple example: a crystal does not explore all its possible configurations because it has privileged directions and it “sticks” to them, see \cite{Strocchi_2005} for a mathematical analysis. However, a more interesting case is given by glasses. Depending on the models, the actual non-ergodicity is in infinite time or due to the relevant time scales, but crucially it corresponds to a variety of states, depending on the paths in the energetic landscape that are taken (or not taken) during the cooling. This can be analyzed as an entropic distance to thermodynamic equilibrium. While this corresponds to a wide variety of “choices”, these various states are very similar and their differences are relatively well described by the introduction of a time dependence for the usual thermodynamical quantities. This corresponds to the so-called “aging dynamics” \cite{Jensen2007}. Glassy dynamics shows
that the absence of a relevant ergodicity is not sufficient in order to obtain phase space changes in our sense, because in this example the various states can be understood in a stable phase space and are not qualitatively different.

As for biology, evolution is both the result of random events at all levels of organization of life and of constraints that canalize it, in particular by excluding, by selection, incompatible random explorations. So, ergodic explorations are also restricted or prevented both by selection and the history of the organism. For example, the presence and the structure of a membrane, or a nucleus, in a cell canalizes also the whole cellular activities along a restricted form of possible dynamics\footnote{See \cite{Machta2011} for an analysis of the taking place of molecular spatial heterogeneity in the membrane, by the coupling of phase transition fluctuations and the cytoskeleton. }.

We find it critical that neither quantum mechanics alone, nor classical physics alone, account for evolution. Both seem to work together. Mutations can be random, acausal, indeterminate quantum events. Yet, they may interfere or happen simultaneously to or be amplified by classical dynamics (see \cite{buiatti2011randomness}). In this amplification, evolution is also not random, as seen in the stunning similarity of the octopus and vertebrate camera eye, independently evolved. Thus evolution is both indeterminate, random and acausal, and yet non random. It is  indeed not sufficiently described by quantum mechanics alone or classical mechanics alone. Life is new.

In other words, classical and quantum randomness superpose and are essential to variability, thus diversity, thus life. None of them is noise. The enthalpic chaotic oscillations of macro-molecules have a classical nature, in physical terms, and are essential to the interaction with DNA and RNA. Quantum randomness of mutation is typically amplified by classical dynamics (including classical randomness), in the interaction between DNA and the proteome, for example. This kind of amplification is necessary in order to understand that changes at the nanometer scale impact the phenotype of the cell or of the organism. Moreover, it is consistent to consider the cell-to-cell interactions and, more generally, ecosystem’s interactions as classical, at least as for their physical aspects, yet affecting the biological observables, jointly with quantum phenomena. Some examples of relevant quantum phenomena are electron tunneling  in cellular respiration \cite{Gray2003}, 
electron transport along DNA  \cite{Winkler2005}, quantum coherence in photosynthesis    \cite{Engel2007,Collini2010}. Moreover, it has been shown that double proton transfer affects spontaneous mutation in RNA
duplexes \cite{Ceron2009}.

When a (random) quantum event at the molecular level (DNA or RNA-DNA or RNA-protein or protein-protein) has consequences at the level of the phenotype, the somatic consequences may persist if they are compatible with the ecosystem and with the ever changing “coherence structure” of the organism as constructed along its history. In particular, it may allow the formation of a new function, organ or tool or different use of an existing tool, thus to the formation of a new properly relevant biological observable. This new observable has at least the same level of unpredictability as the quantum event, but it does not belong to the quantum phase space: it is subject to Darwinian selection at the level of a population, typically, thus it interacts with the ecosystem as such. This is the pertinent level of observability, the space of phenotypes, where biological randomness and unpredictability is now to be analyzed.

Note also that the effects of the classical/quantum blend may show up at a different level of observability and may retroact. First, a mutation or a random difference in the genome, may contribute to the construction of a new phenotype. Second, this phenotype may retroact downwards, to the molecular (or quantum) level. A molecular activity may be excluded, as appearing in cells (organs/organisms) which turn out to be unfit --- selection acts at the level of organisms; methylation and demethylation downwards modify the expression of “genes”. These upwards and downwards activities are part of what we called and discuss MORE below to be the “Kantian whole”, as they contribute to the integration and regulation of and by the whole and the parts. They both contribute to and constrain the biological dynamics and, thus, they do not allow ONE to split the different levels of organization into independent phase spaces.

\section{ Kantian whole and selection}

Before specifying further our approach to biological objects, we have first to challenge the Cartesian and Laplacian view that the fundamental is always elementary and that the elementary is always simple. According to this view, in biology only the molecular analysis would be fundamental. 

Galileo and Einstein proposed fundamental theories of gravitation and inertia, with no references to Democritus’ atoms nor quanta composing their falling bodies or planets. Then, Einstein, and still now physicists, struggle for unification, not reduction of the relativistic field to the quantum one. Boltzmann did not reduce thermodynamics to Newton-Laplace trajectories of particles. He assumed “molecular chaos”, which is far away from the Newton-Laplace mathematical frame of an entailed trajectory in the momentum/position phase space, and unified asymptotically the molecular approach and the II principle of thermodynamics. As a matter of fact, given the ergodic hypothesis, in the thermodynamic integral, an infinite sum, the ratio of particles over a volume stabilizes only at the infinite limit of both. In short, the ergodic hypothesis allows Boltzmann to ignore the entailed newtonian trajectory. The unit of analysis is the volume of each microstate in the phase space. 

Thus, there is no reason in biology to claim that the fundamental must be conceptually elementary (molecular), as this is false also in physics.

Moreover, the proper elementary observable doesn’t need to be simple. “Elementary particles” are not conceptually/mathematically simple, in quantum field theories nor in string theory. In biology, the elementary living component, the cell, is (very) complex, a further anti-Cartesian stand at the core of our proposal: a cell is already a Kantian whole.

As a matter of fact, Kant pointed out that in an “organized being” the parts exist for and by means of the whole, the whole exists for and by means of the parts. The parts perform tasks, typically subsets of their causal consequences, that can be defined only because they are part of a Kantian whole. No reduction to the parts allows understanding of the whole because the relevant degrees of freedom of the parts as associated to the whole are functional (compatibility within the whole and of the whole in the ecosystem) and definable as components of the causal consequences of physical properties of the parts, while the microscopic degrees of freedom of the parts are understood as physical. By this, they include {\it all} the causal consequences of the parts. Further, because of the non-ergodicity of the universe above the level of atoms, where most molecules and organs will never exist, a selective account of the function of a part of a Kantian whole that participates in the continued existence of that whole 
in 
the non-ergodic universe has concrete physical as well as biological meaning. More, in a sense, ergodicity would prevent selection since since it would mean that a negatively selected phenotype would ``come back'' anyway.

As mentioned above and further discussed below, the theoretical frame  establishes the pertinent observables and parameters, i.e. the ever changing and unprestatable phase space of evolution.

In biology, we consider observable and parameters that are derived from or pertinent to Darwinian evolution and this is fundamental for our approach. Their very definition depends on the intended Kantian whole and its integration in AND regulation by an ecosystem. Selection, acting at the level of the evolving Kantian whole in its environment, selects on functions (thus on and by organs in an organism) as interacting with an ecosystem. But the unpresatable task closure achieved by the kantian whole in its niche can only be revealed after the fact by what has succeeded in selection. Thus, the niche itself is co-specified after the selective fact with the kantian whole in a circular way.

Consider for example the crystalline in a vertebrate eye, or the kidney. Both these organs and their functions did not exist before the organisms with crystalline and kidneys were formed. Thus, if we consider the proper biological observable (crystalline, kidney), each phenotypic consequence or set of consequences of a chemical (enzymatic) activity has an a priori  indefinite and unorderable, hence algorithmically undefinable set of potential uses, not pre-definable in the language of physics. Similarly, a membrane bound small protein serving a different function, which by Darwinian pre-adaptation or Gould’s exaptation, may latter become part of the flagellar motor of a bacterium; similarly the double jaws of some vertebrate of 200 million years ago will yield the middle ear bones of today’s vertebrates. There was no mathematical need for the phenotype nor for the function, “listening”, in the physical world. In short, for any single or indefinite set of parts and processes, their causal and quantum 
relations may find some use alone or together, which may allow or increase the capacity of the living being to survive in a new selective environment. That environment may consist of other forms of life as in a mixed ecological community co-evolving with one another often by preadaptations and exaptations. Thus, since the uses of one or many parts or processes alone or together is both indefinite in number and unorderable, we cannot predefine nor, a fortiori, mathematize and altorithmically list those uses ahead of time nor what shall come into existence in the evolving biosphere. In other plain words, we cannot write down equations of motion for these unprestatable, co-constituted, newly relevant observables and parameters in the evolution of the biosphere. As emphasized, by contrast, in physics, laws as equations, dynamics and contour conditions are given in pre-defined phase spaces.

Organisms withstand the intrinsic unstability/unpredictability of the changing phase space, by their relative autonomy. They have an internal, permanently reconstructed autonomy, in Kant’s sense, or Varela’s autopoiesis, that gives them an ever changing, yet “inertial” structural stability. They achieve a closure in a task space by which they reproduce, and evolve and adapt by processes alone or together out of the indefinite and unorderable set of uses, of finding new uses to sustain this in the ongoing evolution of the biosphere. These uses are indefinite as they constitute, when viable, new biological functions and organs, depending on the context. We will formulate this in a physico-mathematical way in section \ref{sec:symchange}. They are unorderable, since variability and diversity manifests itself as the “branching” of evolution, where structure and function of the resulting organs and organisms are on different branches and, by this, uncomparable --- both notions are, of course, epistemic, as we may ”
locally" propose phase spaces and order, whenever this adds to intelligibility of nature.

This situation leads us to introduce the notion of enablement, that we will first define as the role played by a part with respect to the formation of a new observable (mathematically, a new dimension) of the phase space. Examples are given in the next section and we will refine this notion throughout the article.

Let’s summarize this section. In biology selection acts at the level of the whole organism, or what we call a Kantian whole, where as noted above, the function of a part is given by its role in sustaining the whole. Other consequences of the part are side effects. Then selection acting at the level of the Kantian whole, thereby picks out and reveals, after the fact, novel functions and eco-systemic interactions, which are co-constituted by the Kantian whole and its environment as a “task closure” required for reproduction of the Kantian whole in a biological niche. This niche, which is indefinite in features prior to selection revealing what will co-constitute “task closure” for the Kantian whole, allows the tasks’ closure by which an organism survives and reproduces. In contrast to prestated phase spaces in physics, the biologically pertinent phase space is given by the newly relevant observables and the parameters that are needed for the intended analysis, based on life’s proliferation, variation and 
selection. The pertinent observables and parameters, thus the components of the phase space, are the ones that are unprestatble and ever newly relevant for biological functions and interactions. And this is a key point: nothing in biology makes sense, if not analyzed in evolutionary terms. We will summarize the observables as the “phenotype”, that is, as the various (epistemic) components of an organism (organs, functions, interactions \dots ). Thus, it is proliferation, variation and selection grafting novel phenotypes into evolving organisms that reveals, again after the fact, the newly relevant and unprestatable observables and parameters. Thereby, this is our main thesis, the very phase space of evolution changes in unprestatable ways. In consequence, again, we can write no equations of motion for the evolving biosphere, nor know ahead of time the niche boundary conditions so cannot integrate the equations of motion which we do not have. No law entails the evolution of the biosphere. 

\section{ Examples}

Darwinian preadaptations are causal consequences of a part of an organism of no selective significance in the current envionment, which happen to be compatible in a new selective environment where they are thereby selected and so “enabled”. Here we wish to stress the difference between such evolutionary “enablement” for which no entailing law can be found, and entailed trajectories in physics. Then, typically a new biological function comes into existence.

\begin{description}
\item[Example 1.] Swim bladders afford neutral buoyancy in the water column by virtue of the ratio of air to water in the bladder. The preadaptation of the swim bladder derived from the lungs of lung fishes (see \cite{Perry_2001,Zheng_2011} for  more detailed analyses). Water got into some lungs. Now a sac existed with air and water in it, poised to evolve into a swim bladder. With the swim bladder, a new function, neutral buoyancy in the water column, arose. We now ask a fundamental question: Can we name all the possible Darwinian preadaptations, just for humans, in the next three million years? We all say “No”. Why? A first statement of why is that we cannot name all possible selective environments. Nor could we know we had completed the list. Nor could we prestate the one or many features of one or many organisms that might constitute the preadaptation.

Another view of this is the following: once lung fish existed, swim bladders were in the “adjacent possible” of the evolution of the biosphere by Darwinian preadaptations. But we do not know what all the possibilities are for such preadaptation, so do not know the unprestatable and ever changing phase space of evolution. Not only do we not know what “will” happen, we do not even know what “can” happen. 

Yet there is more: Once the swim bladder existed, it thereby constituted a new empty adjacent possible niche. A bacterium or worm might evolve to live only in swim bladders. But before the unprestatable emergence of the swim bladder, we do not know the new empty adjacent possible niche. Yet that niche constitutes the new boundary conditions on natural selection molding the evolution of the bacterium or worm. Again, without laws of motion or knowledge of the boundary conditions, we can have no law entailing the evolution of the biosphere.

This leads to both enablement and radical emergence: The swim bladder was probably selected in a population of fish to function as a swim bladder, but the swim bladder itself as a new adjacent possible empty niche was not selected as a niche, as such. Yet it changes the possible directions of evolution as a worm or bacterium may evolve to live in it. But this means that jointly to but beyond selection, the biosphere creates, by 
variability, recombination and new unpredictable uses, its own 
possibilities of future evolution, that is radical emergence of life from life. Further, both in the sense that quantum acausal processes of mutation may underlie the evolution of the worm, and the deeper fact that no law entails the evolution of the biosphere, the new adjacent possible empty niche does not “cause”, but does “enable” the emergence of a new direction in the evolution of the biosphere.

\item[Example 2.] The bones of the double jaw of some vertebrates evolved into the bones of the middle ears of vertebrates. A new function, hearing, emerged as the “bricolage” of old structures. The same considerations apply to this case of preadaptation as to the evolution of the swim bladder.

\item[Example 3.] The flagella of some bacteria are the “bricolage” of previous components of proteins, which originally had various, unrelated, functions \cite{Liu_2007}. Again, the same considerations apply as those to the evolution of the swim bladder.
\end{description}

\section{Symmetry}

We will understand the historically robust “structure of determination of physics” (which includes unpredictability) by observing that, since Noether and Weyl, physical laws may be described in terms of theoretical symmetries in the intended equations (of the “dynamics”, in a general sense, see below). These symmetries in particular express the fundamental conservation laws of the physical observables (energy, momentum, charges \dots), both in classical and quantum physics. And the conservation properties allow us to compute (they mathematically entail) the trajectories of physical objects as geodetics, by extremizing the pertinent functionals (Hamilton principle applied to the Langrangian functionals), even in Quantum Mechanics, as they allow us to derive the Schroedinger equation.

Note that physics started by first analyzing “trajectories”, from Aristotle to Galileo. Only with the invention of an (analytic) geometry of space (Descartes), could trajectories be placed in a pre-given space, which later became the absolute space of Newtonian laws. As a matter of fact, there is no underlying or background space even in Greek Geometry, which is just a Geometry of figures, handled by rotations and translations (thus, by symmetries). The invention of Descartes spaces was needed for doing both geometry and physics in a pre-stated “background” space. Galileo’s theory provided, by inertia, the conservation law for analyzing the physical invariance, when moving from one Cartesian reference system to another (in modern terms, Galileo’s symmetry group defines the transformations between reference frames and the corresponding invariants).

The invention of the more general notion of “phase space” dates of the late XIX century, when momentum was added to space as an integral component of the analysis of a trajectory (or energy to time, in order to apply the corresponding conservation properties, thus the corresponding theoretical symmetries). In general, the phase spaces are the right spaces of description in the sense that they allow one to specify “trajectories”: if one considers a smaller space, processes would not have a determined trajectory but would be able to behave arbitrarily.

In summary, the historical and conceptual development of physics went as follows:

\begin{itemize}

\item analyze trajectories

\item pull-out the key observables as (relative) invariants (as given by the symmetries)

\item construct out of them the intended phase space.

\end{itemize}

Physical (phase) spaces, thus, are not “already there”, as absolutes underlying phenomena: they are our remarkable and very effective invention in order to make physical phenomena intelligible \cite{weyl1983symmetry,bailly2011}. This approach does not work as for biological observables, we argue.

Thus, the modern work of the physicist begins by setting the phase space and the measure in it, on the grounds of the observables he/she picks out in the “trajectories”, which are meant here in the broadest sense, like in Quantum Physics, where quanta do not go along trajectories, but the wave function does. As for the mathematics, from Descartes’ spaces, invented for accommodating all possible physical trajectories and states, up to the subsequent more general phase spaces, all these spaces are finitistically (axiomatically) describable, because of their symmetries. That is, their regularities, as invariants and invariant preserving transformations in the intended spaces (thus their symmetries), allow a finite description, even if they are infinite or even of infinite dimension.

Consider, say, a tri- (or more) dimensional Cartesian space, since Newton our preferred space. It is infinite, but the three straight lines are given by symmetries (they are axis of rotations) and their right angles as well (right angles, says Euclid, are defined from the most symmetric figure you obtain when crossing two straight lines)\footnote{More generally, modern Category Theory defines Cartesian products in terms of a symmetric commuting diagram. Hilbert and Fock's spaces require a more complex but conceptually similar definition, in terms of invariants and their associated transformations. When adding the different groups of transformations (the symmetries) that allow one to relativize the intended spaces, one obtains the various physical theories that beautifully organize the inert matter, up to today, e.g. Relativity Theory with its group of Poincaré-Lorentz symmetries.}. These symmetries allow us to describe this infinite space in a very synthetic way.

The situation is just slightly more general for a Hilbert space, even of infinite dimension or in other more general cases: mathematics is a science of invariants and invariant preserving transformations, thus of symmetries. These invariants (symmetries) allow us to handle formally infinity in finitely many words. Note that symmetries, in mathematics, have the peculiar status of being both invariant (structural invariants, say) and invariant preserving transformations (as symmetry groups).


In conclusion, in physics the observables, which yield the phase space, derive from the invariants/symmetries in the trajectories. More exactly, they derive from the invariants and the invariant preserving transformations in the intended physical theory. So, Poincaré’s momentum is preserved in the dynamics, similarly as Carnot’s product $pV$ is preserved at constant temperature while $p$ and $V$ may vary. Then one uses these invariants in order to construct the “background space” where the phenomena under analysis can be accommodated. That is, the conceptual construction of the phase space follows the choice of the relevant observables and invariants (symmetries) in the physico-mathematical analysis.

As for biology, let’s go back to our key question: why are we talking of phenotypes, as observables in the biological phase spaces? The main lesson we learn from 20th century physics is that the construction of scientific objectivity (and even of the pertinent objects of science) begins when one gives explicitly the reference system (or phase space) and the metric (the measurement) on it. We do not consider anymore ether or phlogiston or ghosts as pertinent observables nor parameters, thus they have been excluded from our phase spaces. Since Poincaré we learned to add momentum to position and time. Thermodynamics used, as observable forming the phase space, pressure, volume and temperature. Statistical physics considers ensembles of systems. Quantum mechanics takes as state function a probability density in possibly infinite dimensional Hilbert or Fock spaces. As mentioned above, invariants and symmetries allowed these constructions of suitable phase spaces.

Note that in statistical mechanics one may work with a randomly varying number $n$ of particles. Thus, the dimension of the state space stricto sensu, which is $6n$, is not pre-defined. This situation does not, however, lead to relevant difficulties because the possibilities are known (the particles have known nature) and the probabilities of each phase space are given\footnote{In general, $n$ changes either because of chemical reactions, and it is then their rate which is relevant, or because the system is open, in which case the flow of particles is similar to an energetic flow, that is the number of particles plays the same role as energy: they are both fluctuating quantitities obeying conservation laws.}. In other terms, even if the exact finite dimension of the space may be unknown, it has a known probability --- we know the probability it will grow by 1, 2 or more dimensions. The possible extra particles have perfectly known possible properties and states: the pertinent observables and
parameters are known, one just misses: how many? And this becomes a new parameter \dots (see for example \cite{Sethna_2006}).

In quantum mechanics, in order to deal with phase spaces, which are just partially known, a powerful tool, the “density matrix”, has been developed. In such cases, physicists work with the part of the state space that is known and the density matrix takes into account that the system can end up in an unknown region of the state space, by a component called “leakage term”. The point is that this term interferes with the rest of the dynamics in a determined way, which allows us to capture theoretically the situation in spite of the leakage term.

In Quantum Field Theory (QFT) it is even more challenging: particles and anti-particles may be created spontaneously. And so one uses infinite dimensional Hilbert’s spaces and Fock spaces or alike to accommodate them. Of course, quanta are all identical in their different classes: a new electron is an electron \dots they all have the same observable properties. More precisely, the analysis by Feynman diagrams allows us to provide the weight in the quantum state of each possible spontaneous creation and annihilation of particles (and, basically, the more complex a diagram is, the smaller its weight).

In all these cases, the analysis of trajectories or the choice of the object to study (recall the case of the thermodynamic cycle or the law of probability for QM, above) preceded the proposal of the pertinent phase space, which contains the proper observables and parameters for the trajectories of the intended object. Then, as mentioned above, the symmetries of these spaces allowed synthetic, even axiomatic, definitions of these infinite spaces, even of infinite dimension. In other words, the finite description of these possibly infinite spaces, from Descartes to Quantum spaces, is made possible by their regularities: they are given in terms of mathematical symmetries.

\section{ Changing symmetries}\label{sec:symchange}

We will gradually move now towards biology, by continuing though to develop cross references to physics. Our approach is an answer to the physicalist views, often very weakly motivated, but also to many pertinent comments or objections we received from competent colleagues who still prefer to transfer the theoretical approaches to the inert in order to deal also with the living state of the matter.

Since symmetries and the related conservation properties have a major role in physics, it is interesting to consider the paradigmatic case where symmetry changes are at the core of the analysis: critical transitions \cite{Binney}. In these state transitions, “something is not preserved”. In general, this is expressed by the fact that some symmetries are broken or new ones are obtained after the transition (symmetry changes, corresponding to state changes). At the transition, typically, there is the passage to a new “coherence structure” (a non-trivial scale symmetry); mathematically, this is described by the non-analyticity of the pertinent formal development. Consider the classical para-ferromagnetic transition: the system goes from a disordered state to sudden common orientation of spins, up to the complete ordered state of a unique orientation. Or percolation, often based on the formation of fractal structures, that is the iteration of a statistically invariant motif. Similarly for the formation of a snow
flake \dots. In all these circumstances, a “new physical object of observation” is formed. Most of the current analyses deal with transitions at equilibrium; the less studied and more challenging case of far form equilibrium critical transitions may require new mathematical tools, or variants of the powerful renormalization methods. These methods change the pertinent object, yet they are based on symmetries and conservation properties such as energy or other invariants. That is, one obtains a new object, yet not necessarily new observables for the theoretical analysis. Another key mathematical aspect of renormalization is that it analyzes point-wise transitions, that is, mathematically, the physical transition is seen as happening in an isolated mathematical point (isolated with respect to the interval topology, or the topology induced by the usual measurement and the associated metrics).

One can say in full generality that a mathematical frame completely handles the determination of the object it describes as long as no strong enough singularity (i.e. relevant infinity or divergences) shows up to break this very mathematical determination \cite{baillylevel}. In classical statistical fields (at criticality) and in quantum field theories this leads to the necessity of using renormalization methods \cite{fisher1998renormalization,ZinnJustin_2007}. The point of these methods is that when it is impossible to handle mathematically all the interaction of the system in a direct manner (because they lead to infinite quantities and therefore to no relevant account of the situation), one can still analyze parts of the interactions in a systematic manner, typically within arbitrary scale intervals. This allows us to exhibit a symmetry between partial sets of “interactions”, when the arbitrary scales are taken as a parameter.

In this situation, the intelligibility still has an “upward” flavor since renormalization is based on the stability of the equational determination when one considers a part of the interactions occurring in the system. Now, the “locus of the objectivity” is not in the description of the parts but in the stability of the equational determination when taking more and more interactions into account. This is true for critical phenomena, where the parts, atoms for example, can be objectivized outside the system and have a characteristic scale. In general, though, only scale invariance matters and the contingent choice of a fundamental (atomic) scale is irrelevant. Even worse, in quantum fields theories, the parts are not really separable from the whole (this would mean to separate an electron from the field it generates) and there is no relevant elementary scale which would allow ONE to get rid of the infinities (and again this would be quite arbitrary, since the objectivity needs the inter-scale relationship), 
see
for example \cite{ZinnJustin_2007}.

In short, even in physics there are situations where the whole is not the sum of the parts because the parts cannot be summed on (this is not specific to quantum fields and is also relevant for classical fields, in principle). In these situations, the intelligibility is obtained by the scale symmetry which is why fundamental scale choices are arbitrary with respect to this phenomena.

This choice of the object of quantitative and objective analysis is at the core of the scientific enterprise: looking only at molecules as the only pertinent observable of life is worse than reductionist, it is against the history of physics and its audacious unifications and invention of new observables, scale invariances and even conceptual frames.

As for criticality in biology, there exists substantial empirical evidence that living organisms undergo critical transitions \cite{aon2004percolation,Kiyono_2004,GerhardWerner07}. These are mostly analyzed as limit situations, either never really reached by an organism or as occasional point-wise transitions. Or also, as researchers nicely claim in specific analysis: a biological system, a cell genetic regulatory networks, brain and brain slices \dots are “poised at criticality” \cite{nykter2008gene, mora2010biological}. In other words, critical state transitions happen continually.

Thus, as for the pertinent observables, the phenotypes, we propose to understand evolutionary trajectories as cascades of critical transitions, thus of symmetry changes. In this perspective, one cannot pre-give, nor formally pre-define, the phase space for the biological dynamics, in contrast to what has been done for the profound mathematical frame for physics. This does not forbid a scientific analysis of life. This may just be given in different terms. We currently work, for example, at the mathematical analysis of these cascades of symmetry changes in embrogenesis and their role as for the variability of embryogenetic trajectories.  Below we propose a positive science that analyses the worlds Kantian wholes mutually make.

As for evolution, there is no possible equational entailment nor a causal structure of determination derived from such entailment, as in physics. the point is that these are better understood and correlated, since the work of Noether and Weyl in the last century, as symmetries in the intended equations, where they express the underlying invariants and invariant preserving transformations. No theoretical symmetries, no equations, thus no laws and no entailed causes allow the mathematical deduction of biological trajectories in pre-given phase spaces --- at least not in the deep and strong sense established by the physico-mathematical theories. Observe that the robust, clear, powerful physico-mathematical sense of entailing law has been permeating all sciences, including societal ones, economics among others. If we are correct, this permeating physico-mathematical sense of entailing law must be given up for unentailed diachronic evolution in biology, in economic evolution, and cultural evolution.

As a fundamental example of symmetry change, observe that mitosis yields different proteome distributions, differences in DNA or DNA expressions,
in membranes or organelles: the symmetries are not preserved. In a multi-cellular organism, each mitosis asymmetrically reconstructs a new coherent “Kantian whole”, in the sense of the physics of critical transitions: a new tissue matrix, new collagen structure, new cell-to-cell connections \dots. And we undergo millions of mitosis each minute. More, this is not “noise”: this is variability, which yields diversity, which is at the core of evolution and even of stability of an organism or an ecosystem. Organisms and ecosystems are structurally stable, also because they are Kantian wholes that permanently and non-identically reconstruct themselves: they do it in an always different, thus adaptive, way. They change the coherence structure, thus its symmetries. This reconstruction is thus random, but also not random, as it heavily depends on \emph{constraints}, such as the proteins types imposed by the DNA, the relative geometric distribution of cells in embryogenesis, interactions in an organism, in a niche, but
also on the opposite of constraints, the autonomy of Kantian wholes.

In the interaction with the ecosystem, the evolutionary trajectory of an organism is characterized by the co-constitution of new interfaces, i.e. new functions and organs that are the proper observables for the Darwinian analysis. And the change of a (major) function induces a change in the global Kantian whole as a coherence structure, that is it changes the internal symmetries: the fish with the new bladder will swim differently, its heart-vascular system will relevantly change \dots.

Organisms transform the ecosystem while transforming themselves and they can stand/do it because they have an internal preserved universe (Bernard’s ”milieu interior"). Its stability is maintained also by slightly, yet constantly changing internal symmetries. The notion of extended criticality in biology (life is always in an extended critical transition with respect to all the pertinent parameters, see \cite{bailly2011,longo2011c}) focuses on the dynamics of symmetry changes and provides an insight into the permanent, ontogenetic and evolutionary adaptability, as long as these changes are compatible with the co-constituted Kantian whole and the ecosystem. As we said, autonomy is integrated in and regulated by constraints, with an organism itself and of an organism within an ecosystem. Autonomy makes no sense without constraints and constraints apply to an autonomous Kantian whole. So constraints shape autonomy, which in turn modifies constraints, within the margin of viability, i. e. within the limits of
the interval of extended criticality. The extended critical transition proper to the biological dynamics does not allow one to prestate the symmetries and the correlated phase space.

Consider, say, a microbial ecosystem in a human. It has some 150 different microbial species in the intestinal tract. Each person’s ecosystem is unique, and tends largely to be restored following antibiotic treatment. Each of these microbes is a Kantian whole, and in ways we do not understand yet, the “community” in the intestines co-creates their worlds together, co-creating the niches by which each and all achieve, with the surrounding human tissue, a task closure that is “always” sustained even if it may change by immigration of new microbial species into the community and extinction of old species in the community. With such community membership turnover, or community assembly, the phase space of the system is undergoing continual and open ended changes. Moreover, given the rate of mutation in microbial populations, it is very likely that these microbial communities are also co-evolving with one another on a rapid time scale. Again, the phase space is continually changing as are the symmetries.

Can one have a complete description of actual and potential biological niches? If so, the description seems to be incompressible, in the sense that any linguistic description may require new names and meanings for the new unprestable functions, where functions and their names make only sense in the newly co-constructed biological and historical (linguistic) environment. Even for existing niches, short descriptions are given from a specific perspective (they are very epistemic), looking at a purpose, say. One finds out a feature in a niche, because you observe that if it goes away the intended organisms dies. In other terms, niches are compared by differences: one may not be able to prove that two niches are identical or equivalent (in supporting life), but one may show that two niches are different. Once more, there are no symmetries organizing over time these spaces and their internal relations. Mathematically, no symmetry (groups) nor (partial-)order (semigroups) organize the phase spaces of phenotypes, in 
contrast to physical phase spaces.

In order to summarize these preliminary remarks, it should be clear that our aim is to show that the powerful methods of physics that allowed us to pre-define phase spaces on the grounds of the observables and the invariants in the ''trajectories" (the symmetries in the equations) do not apply in biology.

Finally, here is one of the many logical challenges posed by evolution: the circularity {\it of} the definition of niches is more than the circularity {\it in} the definitions. The “in the definitions” circularity concerns the quantities (or quantitative distributions) of given observables. Typically, a numerical function defined by recursion or by impredicative tools yields a circularity in the definition and poses no mathematical nor logical problems, in contemporary logic (this is so also for recursive definitions of metabolic cycles in biology, see \cite{Mossio2009489}. Similarly, a river flow, which shapes its own border, presents technical difficulties for a careful equational description of its dynamics, but no mathematical nor logical impossibility: one has to optimize a highly non linear and large action/reaction system, yielding a dynamically constructed geodetic, the river path, in perfectly known phase spaces (momentum and space or energy and time, say, as pertinent observables and variables).

The circularity “of the definitions” applies, instead, when it is impossible to prestate the phase space, so the very novel interaction (including the “boundary conditions” in the niche and the biological dynamics) co-defines new observables. The circularity then radically differs from the one in the definition, since it is at the meta-theoretical (meta-linguistic) level: which are the observables and variables to put in the equations? It is not just within prestatable yet circular equations within the theory (ordinary recursion and extended non---linear dynamics), but in the ever changing observables, the phenotypes and the biological functions in a circularly co-specified niche. From this follows our statement that, mathematically and logically, no law entails the evolution of the biosphere.

\section{ Enablement, causality, and randomness}


\subsection{ The Terminus of a Physics Worldview at the Watershed of Life}

The instability of biological theoretical symmetries is not, of course, the end of science, but it is the terminus of the flat transfer of physico-mathematical methods of physics, taught us from Newton onward, to biology. In biological evolution we cannot use the same very rich interaction with mathematics at the core of physical theories. Consequently, our analysis, if correct, puts an end, a fortiori, to any form of reductionism, either to existing physical theories or to the even simpler theories of information, either Shannon or Turing, both embedded in the fragments of current mathematics dealing with discrete structures and data types, whose phase spaces are even more strictly pre-given. Because the ever changing phase space of biological evolution cannot be prestated and its sample space is not statable, one cannot compute “the entropy of the Source”, a la Shannon, nor can one define the algorithmic generation of a sequence of prestated symbols, a la Turing.

The scientific answer we propose to this end of the physicalist certitudes, is based on an analysis of the notion of “enablement” in evolution (and ontogenesis, to be discussed elsewhere). In turn the enablement concerns how Kantian wholes co-create their worlds such that they can exist in the non-ergodic universe above the level of atoms.

Our thesis then is that evolution as a “diachronic process” of becoming (but ontogenesis as well) just “enables”, but does not cause, the forthcoming state of affairs, in the sense specified below. Our point of view reflects the fact that Newton’s entailed trajectories mathematized Aristotle’s “efficient cause”. In short, in our view, such \emph{entailed causal relations must be replaced by “enablement” relations}, in evolutionary biology.

Typically, a niche enables the survival of an otherwise incompatible/impossible form of life, it does not cause it. At most, a cause may be found in the ''difference" (a mutation, say) that induced the phenotypic variation at stake, as spelled out below.

This new perspective will be motivated, on one side, by the understanding of physical “causes and determinations” in terms of symmetries, along the lines above of modern physics; on the other, by an analysis of biological “trajectories” in phylogenesis (and ontogenesis), as continual symmetry changes. We claim that without invariances and symmetries, thus (possibly equational) laws, entailed “causes” cannot be defined. As part of this understanding, we will discuss unentailed causal relations in a restricted sense, in terms of “differential causes”.

In summary, in biology, there is no way to extract the pertinent biological observables as invariant properties, preserved by mathematical symmetries, and then transfer these observables to a “background phase space”, as physicists did, since (Galileo and) Newton, by using first Descartes’ spaces, then by inventing more general phase spaces (e. g. Hilbert spaces). In short, life, unentailed, “bubbles forth” (as Heraclutus said 2500 years ago) and organisms (their phenotypes) co-constitute their own phase space. Kantian Autonomy and constraints and affordances of the environment are the two faces of the same medal, dynamically shaping each other. As remarked in the Overview, an organism exists as a Kantian whole by virtue of a closure in some set of “tasks” by which it reproduces. But these tasks can typically only be completed via the biotic and abiotic environment. In turn, in evolving Kantian wholes under natural selection acting on those wholes, only after the selective fact are the relevant features of 
the tasks, their closure, and the niche revealed. Thus the tasks and the niche are circularly co-constituted by selection acting on the evolving population of organisms. The Kantian whole with its task closure and its niche are both not separately specifiable, nor prestatable prior to the results of selection.

\subsection{Causes and Enablement}\label{subsec:cause}

Since symmetries are unstable, causality in biology cannot be understood as “entailed causality” as in physics and this will lead us to the proposal that “causal relations are only differential causes”. If a bacterium causes pneumonia, or a mutation causes one of the rare monogenetic diseases (anemia falciformis, say), this is a cause and it is differential, i. e. it is a difference with respect to what is fairly considered "normal" and causes an anormality in the phenotype.

A classical mistake is to say: this mutation causes an idiot child (a famous genetic disorder, phenylketonuria), thus \dots the gene affected by the mutation is the gene of intelligence, or: here is the gene that causes/determines the intelligence. In logical terms, it consists in deducing from “notA implies notB”, that “A implies B”: an amazing mistake. All that we know is a causal correlation of differences.

We then propose to consider things differently. The single observed or induced difference, a mutation with a somatic effect, a stone bumping on someone’s head, or a carcinogen (asbestos) does cause a problem; that is the causal dictionary is suitable to describe a differential cause-effect relation. The differential cause modifies the space of possibilites, that is the compatibility of the organism with the ecosystem. In other terms, it modifies the ''enablement relations". In some case, the modified frame may be viable.

In other words, the differential causes modify the coherence structure of an organism, a niche, an ecosystem. So enablement is modified: a niche may be no longer suitable for an organism. Either selection may exclude the modified organism or a change in a niche, due to a differential physical cause (a climate change, say), may negatively select existing organisms or act on them differently, since the enablement relations differ.

Yet, by the following two arguments, we claim that only the differential relations may be soundly considered causal.

\begin{enumerate}
\item  In physics, in presence of an explicit equational determination, causes are given by a formal symmetry breaking in equations. Typically, $f = ma$, a symmetric relation, means, for Newton, that a force causes an acceleration, asymmetrically. In a synthetic/naive way, one may say that Einstein reverses the causal implication, as a space curvature "causes" an acceleration that "causes" a field, thus a force. One may consider the application of a Newtonian force as a differential cause, as the inertial movement is the "default" state (”nothing happens” if no force is applied). This analysis cannot be transferred to biology, inasmuch symmetries are not stable and, thus, one cannot write equations for phylogenetic trajectories (nor break their symmetries). Moreover, the default state is far form being inertia (see below).

\item As just mentioned the default state in physics is inertia. In biology instead, the default state is ”activity". In short, following an idea to be developed elsewhere, the default state is the \emph{never identical iteration of a morphogenetic process}, by \emph{proliferation} and \emph{mobility} (see also \cite{Society}). In this paper’s terms, the default state involves continual critical transitions, thus symmetry changes, thus phase space changes. As a consequence, an organism, a species, does not need a cause to be active, e. g. to proliferate and occupy a new niche. It only needs to be a Kantian whole enabled to survive by changing.
\end{enumerate}

How do biological niches relate to physical boundary conditions? Consider a billiard table with its four edges which are the boundary condition on the motion of a billiard ball. The boundary conditions play a causal role in the motion of the ball, according to Newton's third law of motion. Hence integration of the laws of motion in differential equations with initial and boundary conditions do yield the future and past trajectories of the ball. Energy as a conserved observable allows us to mathematically describe the dynamics. Consider then an adjacent possible empty niche, say the swim bladder. Is it a boundary condition?
Not in the sense as in physics, since the swim bladder enables the worm or the bacterium to live and evolve, according to {\it un-prestatable enabling relations}: the features of the swim bladder to be used by the new organism to achieve task closure in its environment may be radically new. In contrast, in physics, the energy conservation properties allow us to derive the equations of the action/reaction system \dots proper to the physical phenomenon; all the pertinent observables and symmetries, in a billiard and balls system, are perfectly known, or completely pre-given.

In other words, we cannot know ahead of time the co-constituted observables (functions, typically) of the worm and swim bladder that allow the worm to live in the swim bladder. Hence we cannot, even knowing the existence of the swim bladder, state what features of the swim bladder as a niche actually constitute the "actual niche", or even the ever changing niche as the worm evolves.

In general, the organism adjusting to a new niche may be a "hopeful monster", in the sense of Goldschmidt, that is the result of a "pathology" \cite{Dietrich,Gould1977}.

The notions of “normal” and “pathological” makes no sense in physics, where all causes may be considered “differential”. In a sense, these differing notions allowed us to distinguish between enablement and causality, due to a difference in the normal web of interactions. In evolution, a difference (a mutation) may cause a “pathology”, or a hopeful monster. But this monster, if not killed by selection, may be enabled to survive by and in a co-constituted niche.

Thus, besides the centrality of enablement, we may maintain the notion of cause --- and it would be a mistake to exclude it from the biological dictionary. As a matter of fact, one goes to the doctor and rightly asks for the cause of pneumonia --- not only what enabled it: find and kill the bacterium, please, which is the cause. Yet, that bacterium has been enabled to grow excessively by a weak lung, a defective immune system or bad habits \dots so, the therapy should not stop at this differential cause, but investigate enablement as well \cite{denis2009}. And good doctors do it, without necessarily naming it.

\subsection{Randomness}
 
Since Descartes and Galileo’s invention of modern physical space, both predictable and unpredictable (= random) events are considered within the pre-determined Cartesian space of all possible trajectories and phenomena. Newton’s laws are stated in this pre-given and stable mathematical space. Of course, hazards (= random events) are possible, but they occur within a perfectly pre-given space of possibilities: like throwing dice --- it is a risk, but within the six possible outputs, no more, no less. This is why betting on dices is rational: one can compute the probabilities and evaluate the risk. That is, randomness is “measurable” and its measure is given by probability theory, since Pascal and Laplace, a theory of measure in pre-given spaces of possibilities (the pertinent phase spaces).
 
More precisely, the measure (the probabilities) is given in terms of (relative) probabilities defined by symmetries with respect to the observable in a prestated phase space. For the dice, the probabilities of the 6 outcomes correspond to the geometrical symmetries of the object. A more sophisticated example is the microcanonical ensemble of statistical mechanics, where the microstates with the same energy have the same probability, on the grounds of the ergodic hypothesis.
 
As a further consequence, we will hint at the need of a different concept of randomness in biology, since we cannot apply a normal probability measure to biological unpredictability: in absence of a pre-given space of possibilities, biological randomness cannot be measured by probabilities, which are only definable on a prestated phase space or sample space. This is yet another crucial difference with respect to the various notions of randomness in physics, which are all measurable,
 
Randomness enhances variability and diversity. It is thus at the core of evolution: it permanently gives diverging evolutionary paths, as theoretical bifurcations in the formation of phenotypes. And note that variability and diversity are also a component of the structural stability of organisms, species and ecosystems, as Kantian wholes alone and together. Differentiation and diversity within an organism, a species and an ecosystem contribute to their viability and robustness.
 
Note finally, and this is crucial, that mutations, as one of the primary (differential) causes of the phylogenetic drift underlying evolution, must be understood in terms of a "non-conservation principle'' of biological observables. This is the exact opposite of the symmetries and conservation properties that govern physics and the related equational and causal approaches. There is of course structural stability, in biology, which means similar, but never identical iteration of phenotypes. Yet, evolution requires also and intrinsically a non-conservation principle in order to be made intelligible.

Moreover, major phenomena of convergent evolution shape similarly organs and organisms. For example, as noted above, the convergent evolution of the octopus and vertebrate eye may also follow, on one side, random quantum based acausal and indeterminate mutations, but, on the other, it is also ''non-random" as both eyes converge to analogous biological structures, probably due to physical and biological similar constraints --- acting as co-constituted borders or as selection. The convergent evolution of marsupial and mammalian forms, like the Tasmanian wolf and mammalian wolf are other examples of convergent, non-random components of evolution, in the limited sense above.
 
In these situations with a superposition of quantum and classical physics, the remarkable coexistence of indeterminate acausal quantum mutations and non-random convergence to the camera eye, hence quantum random acausal, indeterminate yet also non-random, comes from the fact that we are not considering a stable phase space: these examples are not similar to a dice which falls two times on the same side (or two quantum measurements that yield the same result). On the contrary, we have two biological trajectories which end up constituting similar symmetries and a similar function. And this happens without a prestated space, with stable underlying symmetries, which would allow us to provide a measure of the chance for it to happen.
 
Conversely, evolution may give divergent answers to the same or to similar physical constraints. That is, the same function, moving for example, or breathing, may be biologically implemented in very different ways. For example, tracheas in insects versus vertebrates’ lungs (combined with the vascular system), due both to different contexts (different biological internal and external constraints) and random symmetry changes in evolutionary paths.
 
More generally, conservation or optimality properties of physical observables (the various forms of physical energy, for example) cannot help to determine the evolutionary trajectory of an organism. No principle of “least free energy” (or “least time consumption of free energy”, if it applies) can help to predict or understand the formation of a proper biological observable, the organism as object of selection. Only locally, i. e. the form of some organs, where exchange of matter or energy dominates (lungs, vascular system, phyllotaxis \dots), may be shaped by the dominating physical forces and, thus, their forms follow causally optimality principles (fractal structures or alike, typically).
 
Yet, very different biological answers to the “same” physical context make phenotypes incomparable, in terms of physical optima: production of energy or even exchanging oxygen may be dealt with in very different ways, by organisms in the “same” ecosystem. Moreover, in terms of physical or also biological observables, the front legs of an elephant are not better nor worst than those of a Kangaroo: front podia of tetrapodes diverged (broke symmetries differently) in different biological niches and with respect to different internal milieu. And none of the issuing paths is “better” than the other, nor followed physical optimality criteria, even less biological ones: each is just a possible variation on an original common theme, just compatible with the co-constituted ecosystem that enabled them. In general, there is no way to define a real valued (Lagrangian) functional to be extremized as for biological functions, as this would require an ordered space (a real valued functional), where “this biological 
function (or 
phenotype)” could 
be said to be “better” than “that biological function (or phenotype)”.
 
Note that this un-comparability (or absence of even a partial order between phenotypes), is also due to the relative/partial independence of niches. These are co-constituted by organisms. Niches enable what evolves.

\section{Conclusion}
 
We stressed the role of invariance, symmetries and conservation properties in physical theories. In biology, symmetries at the phenotypic level, are continually changed, beginning with the least mitosis, up to the “structural bifurcations” which yield speciations in evolution. Thus, there are no biological symmetries that are a priori preserved, except and for some time, some basic structures such as bauplans (still more or less deeply modified during evolution). There are no sufficiently stable mathematical regularities and transformations, to allow an equational and law like description entailing the phylogenetic and ontogenetic trajectories. These are cascades of symmetry changes and thus just cumulative historical dynamics. And each symmetry change is associated to a random event (quantum or classical), at least for the breaking of symmetries, while the global shaping of the trajectory, by selection say, is also due to non-random events. In this sense biological trajectories are generic, that is just 
possible 
ones, and yield a historical result, an individuated, specific organism (see \cite{bailly2011,longo2011c}).
 
As a consequence, this sum of individual and individualizing histories, co-constitued within an ever changing ecosystem, does not allow a compressed, finite formal description of the space of possibilities, an actual biological phase space (functions, phenotypes, organisms): these possibilities are each the result of an unpredictable sequence of symmetry breakings, in contrast to the invariant (conservation) properties which characterize physical “trajectories”, in the broad sense (extended to Hilbert's spaces, in Quantum Mechanics). The physical phase spaces have been mathematically designed, in terms of symmetries, exactly in order to accommodate the physical invariants of trajectories. Recall the interplay between Cartesian Spaces and Galileo’s inertia, at the foundation of modern physics and its phase spaces; similarly for Poincaré-Lorentz groups in Relativity Theory or Hilbert and Fock’s spaces in QM.
 
In this sense, there are no laws that entail, as in physics, the becoming of the biosphere, and a fortiori, the econosphere, or culture or history, or life in general. In the same sense, geodetic principle mathematically forces physical objects never to go wrong. A falling stone follows exactly the gravitational arrow. A river goes along the shortest path to the sea, it may adjust it by nonlinear well definable interactions as mentioned above, but it will never go wrong. These are all geodetics.  Living entities, instead, go wrong most of the time: most organisms are extinct, the majority of fecundations, in mammals, do not lead to a birth, an amoeba does not follows, exactly, a curving gradient --- by retention it would first go along the tangent, then correct the trajectory, in a protensive action. In short, life goes wrong most of the time, but it “adjusts” to the environment and changes the environment, if possible. It maintains itself, always within an extend critical interval, whose limits are the 
edge of death, by changing the observables, the phenotypes, that is the very nature of the living object.
 
 If we are right in our conclusion, the strong reductionist dream of a theory that entails the full becoming of the universe is wrong. With life, we reach the end of a physics worldview that has dominated us since Newton. Yet the biosphere is magnificant in its enabled, radically emergent becoming, the most complex system we know of in the universe. If no law entails this becoming, then our belief that such entailing law is a necessary condition for the emergence of highly organized integrated complex systems is wrong. As Darwin said, “There is a grandeur in this view of life\dots”, now seen to be beyond entailing laws. Then we must ask new scientific questions, for this co-constitution by Kantian wholes as they co-evolve and make their worlds together must be a central part of how the biosphere emerges. We do not begin to understand this co-constitution and enabled radical emergent becoming in which the biosphere, jointly to but beyond selection, creates its own future possibilities of becoming. So too does 
the 
evolving econosphere and culture.
 
 Kantian wholes making worlds with one another in the biosphere, our approach, is part of a positive science, beyond entailing laws, for why the biosphere is the most complex system we know of in the universe and works so well. And more, we may be able to analyze the growth of the adjacent possible as the ever new niches enabled by the Kantian wholes living together. "Enablement" allows one or more Kantian wholes to co-create a world and environment that allows them to exist in the non-ergodic universe where most complex things won't exist. Yes, life is  caught in a causal web, but lives also in a web of enablement and radical emergence.
 
Two thousand five hundred years ago Heraclitus said, “The world bubbles forth”. He may not be correct for physics. But at the watershed of life, we agree: Life bubbles forth.

\paragraph{Acknowledgments:} Stuart Kauffman is a Distinguished Finnish Professor supported by the Tekes Foundation, Finland.
\bibliographystyle{alpha}
\bibliography{bib}

\end{document}